\documentclass[12pt]{emulateapj}

\def\lsim{\lower.5ex\hbox{$\; \buildrel < \over \sim \;$}}
\def\gsim{\lower.5ex\hbox{$\; \buildrel > \over \sim \;$}}

\newcommand{\msun}{$M_\odot$}
\newcommand{\mearth}{$M_\earth$}

\shorttitle{A runaway WR star as the origin of $^{26}$Al in the early solar system} \shortauthors{Tatischeff et al.}

\begin{document}

\title{A runaway Wolf-Rayet star as the origin of $^{26}$Al in the early solar system}

\author{Vincent Tatischeff, Jean Duprat}
\affil{Centre de Spectrom\'etrie Nucl\'eaire et de Spectrom\'etrie de Masse, 
IN2P3-CNRS and Univ Paris-Sud, F-91405 Orsay Cedex, France}
\and
\author{Nicolas de S\'er\'eville}
\affil{Institut de Physique Nucl\'eaire d'Orsay, 
IN2P3-CNRS and Univ Paris-Sud, F-91405 Orsay Cedex, France}

\begin{abstract}
Establishing the origin of the short-lived radionuclide (SLR) $^{26}$Al, which was present in refractory inclusions in primitive meteorites, has profound implications for the astrophysical context of solar system formation. Recent observations that $^{26}$Al was homogeneously distributed in the inner solar system prove that this SLR has a stellar origin. In this Letter, we address the issue of the incorporation of hot $^{26}$Al-rich stellar ejecta into the cold protosolar nebula. We first show that the $^{26}$Al atoms produced by a population of massive stars in an OB association cannot be injected into protostellar cores with enough efficiency. We then show that this SLR likely originated in a Wolf-Rayet star that escaped from its parent cluster and interacted with a neighboring molecular cloud. The explosion of this runaway star as a supernova probably triggered the formation of the solar system. This scenario also accounts for the meteoritic abundance of $^{41}$Ca.

\end{abstract}


\section{Introduction}

More than 30 years after the discovery of \citet{lee76} that calcium-aluminum-rich inclusions (CAIs) from the Allende meteorite contained $^{26}$Al (mean lifetime $\tau_{26}=1.03\times10^6$~yr), the origin of this short-lived radionuclide (SLR) remains an open question. High precision Mg isotopic analyses of asteroids and bulk rocks from terrestrial planets \citep{thr06}, as well as recent micrometer-scale data in chondrules \citep{vil09}, showed that $^{26}$Al was homogenously distributed over at least the inner part of the solar system, i.e., over a reservoir of mass $>2$~\mearth, and that no significant amount of freshly made $^{26}$Al was added to the protoplanetary disk after the CAI formation. The maximum amount of $^{26}$Al that could have been synthesized by in situ particle irradiation during the short duration of CAI formation \citep[$\sim 10^5$~yr;][]{biz04} can account for the canonical $^{26}$Al/$^{27}{\rm Al}=5\times 10^{-5}$ over a rocky reservoir of only $\sim$0.1~\mearth~\citep{dup07}. Thus, the origin of this SLR cannot be related to the nonthermal activity of the young Sun and has to be searched for in a stellar nucleosynthetic event contemporary with the formation of the solar system.

An origin of SLRs in an asymptotic giant branch (AGB) star has been proposed \citep{was94,tri09}, but AGB stars are not associated with star-forming regions and the probability of a chance encounter between an AGB star and a star-forming molecular cloud is very low \citep{kas94}. It is more likely that the protosolar system was contaminated  by material freshly ejected from a massive star, either a Type II SN \citep[e.g.][]{cam77} or a Wolf-Rayet (WR) star \citep[e.g.][]{arn97}. Massive stars have a profound influence on the surrounding molecular clouds and the process of star formation \citep[e.g.][]{lee07}. \citet{cam77} first suggested that a supernova (SN) responsible for injecting SLRs into the presolar nebula may also have been responsible for triggering the formation of the solar system. Detailed numerical simulations have shown that such simultaneous triggering and injection is possible, but the injection efficiency is lower than required \citep{bos10}. Alternatively, it has been suggested that a nearby SN ($\sim$0.3~pc) may have injected SLRs into the already-formed protoplanetary disk of the solar system \citep[see][and references therein]{oue07}. In this scenario, it is assumed that the Sun was born in a large stellar cluster containing massive stars. But this model is questionable, because (1) protoplanetary disks in the vicinity of massive stars are exposed to a rapid photoevaporation and (2) the main-sequence lifetime of even the most massive stars is too long as compared to the mean lifetime of protoplanetary disks \citep{gou08}.

\citet{gou09} and \citet{gai09} recently suggested that the Sun is born in a stellar cluster of second generation, whose formation was triggered by the activity of a neighboring OB association. There are many observations of OB associations divided in spatially separated subgroups of different ages \citep[e.g.][]{bla64}, as well as observations of young stellar objects located on the border of H~II regions \citep[e.g.][]{kar09} and superbubbles \citep[e.g.][]{lee09}. Adopting such an astrophysical context, we study in this Letter how hot stellar debris enriched in $^{26}$Al could be injected into a cold protostellar nebula. We show in Section~2 that $^{26}$Al produced by a population of massive stars in an OB association may not be delivered into molecular cores efficiently enough. We then study in Section~3 a possibility already mentioned in the pioneering work of \citet{arn97} and more recently by \citet{gai09} that the presolar nebula was contaminated by $^{26}$Al produced by a WR star that escaped from its parent cluster.

\section{$^{26}$Al production by an OB association in a superbubble}

Most massive stars are born in OB associations, where multiple stellar winds merge and expand to form large cavities of hot gas known as superbubbles \citep[see, e.g.,][]{par04}. The subsequent SNe generally explode inside the wind-generated superbubble. The radius of a superbubble can be estimated from the standard wind bubble theory \citep{wea77}:
\begin{equation}
R_{\rm SB} \simeq (22~{\rm pc}) t_{\rm Myr}^{3/5} N_{*,30}^{1/5} n_{\rm H,100}^{-1/5}~,
\label{eq1}
\end{equation}
where $t_{\rm Myr}$ is the time in units of Myr after the onset of massive star formation (assumed to be coeval for all stars), $N_{*,30}=N_*/30$ where $N_*$ is the number of massive stars in the 8--120~\msun~mass range, and $n_{\rm H,100}=n_{\rm H}/(100~{\rm cm^{-3}})$ where $n_{\rm H}$ is the mean H number density in the ambient interstellar medium. The superbubble radius is generally given as a function of the stellar wind mechanical power, $L_w$, instead of the number of massive stars \citep[e.g.][]{mac88}. But Equation~(\ref{eq1}) uses the recent result of \citet{vos09} that the mean wind power per star from a coeval population of massive stars is nearly constant with time for $\sim$5~Myr and amounts to $\approx$$1.5 \times 10^{36}$~erg~s$^{-1}$. Similarly, the characteristic temperature and H number density in the interior of a superbubble can be written as \citep{wea77,mac88}
\begin{eqnarray}
T_{\rm SB} & \simeq & (5.7 \times 10^6~{\rm K}) t_{\rm Myr}^{-6/35} N_{*,30}^{8/35} n_{\rm H,100}^{2/35}~, \label{eq2}
\\
n_{\rm SB} & \simeq & (0.17~{\rm cm^{-3}}) t_{\rm Myr}^{-22/35} N_{*,30}^{6/35} n_{\rm H,100}^{19/35}~. 
\label{eq3}
\end{eqnarray}

\begin{figure}
\includegraphics[scale=.44]{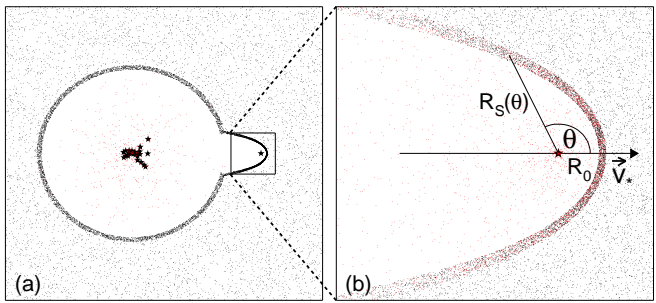}
\figcaption{Sketch illustrating the ejection of $^{26}$Al by (a) a population of massive stars in an OB association and (b) a WR star ``running away'' from its parent cluster.
\label{fig1}}
\end{figure}

WR wind and SN ejections of $^{26}$Al occur at $t_{\rm Myr} \gsim 3$ \citep{vos09}, when the superbubble blown by the winds from the main-sequence stars has already reached a radius of several tens of pc (Equation~(\ref{eq1})). Noteworthy, SN blast waves within a superbubble will usually become subsonic in the hot gas before they reach the supershell of swept-up interstellar material \citep{mac88,par04}. This is true as well for winds of WR stars. Thus, most nuclei synthesized in massive stars first thermalize in the hot superbubble interior. Further incorporation of this material into molecular clouds and star-forming systems takes more than 10~Myr \citep{mey00}, by which time the $^{26}$Al will have decayed. 

To solve this issue, \citet{gai09} proposed that $^{26}$Al ejected in WR winds can be rapidly incorporated into high-speed ($\sim$1000~km~s$^{-1}$) refractory dust grains of $\sim$0.01--0.1~$\mu$m size, that could dynamically decouple from the shocked wind gas and imbed themselves into the surrounding molecular material. But this proposal has two shortcomings. First, WR stars are thought to be a major contributor to the Galactic $^{26}$Al detected through its gamma-ray decay line at $E_\gamma=1809$~keV, and high-resolution spectroscopic observations of this emission with the {\it RHESSI} and {\it INTEGRAL} gamma-ray satellites have shown that the line is narrow, $\Delta E_\gamma=1$--2~keV FWHM, consistent with the instrumental resolution \citep[see][and references therein]{die06}. The non-detection of Doppler broadening in the Galactic 1809-keV line provides an upper limit on the mean velocity of the emitting $^{26}$Al nuclei: $v_{\rm max} \sim 0.5 c \Delta E_\gamma / E_\gamma \sim 150$~km~s$^{-1}$ (here, $c$ is the speed of light). This maximum velocity is much lower than the speed that dust grains must acquire to survive sputtering as they pass the WR wind termination shock \citep{gai09}. Secondly, most grains formed in WR winds will slow down and stop in the superbubble interior before reaching the supershell. According to the classical estimate of \cite{spi78}, the range of a grain of size $a_{\rm gr}$ and typical density $\rho_{\rm gr} \sim 2~$g~cm$^{-3}$ is $X_{\rm gr}=a_{\rm gr}\rho_{\rm gr}=(2 \times 10^{-6}~{\rm g~cm^{-2}})(a_{\rm gr}/0.01~{\rm \mu m})$. In comparison, the radial path length in a superbubble is 
\begin{eqnarray}
X_{\rm SB} & = & 1.4 m_{\rm H} \int_{0}^{R_{\rm SB}} n(r) dr \nonumber \\
& \simeq & (4.6 \times 10^{-5}~{\rm g~cm^{-2}}) t_{\rm Myr}^{-1/35} N_{*,30}^{13/35} n_{\rm H,100}^{12/35}~,
\label{eq4}
\end{eqnarray}
where $m_{\rm H}$ is the H mass and $n(r)=n_{\rm SB}[1-(r/R_{\rm SB})]^{-2/5}$ \citep{wea77}. Thus, grains with $a_{\rm gr} \lsim 0.2~\mu$m do not reach the supershell. In fact, even much larger grains should stop in the superbubble interior, because the Spitzer formula can largely overestimate the range of interstellar dust grains in hot plasmas \citep{rag02}. 

Dense clumps of molecular gas can be engulfed by the growing superbubble, if they were not swept up by the expanding supershell \citep[e.g.][]{par04}. These clumps could potentially be enriched in $^{26}$Al synthesized by WR stars and Type II SNe in the OB association. But recent two-dimensional  hydrodynamic simulations \citep{bos08,bos10} suggest that the amount of $^{26}$Al that could be injected into such a molecular cloud core would be too low to explain the solar system's canonical $^{26}$Al/$^{27}$Al ratio. Boss et al. found that only 2--5$\times 10^{-5}$~\msun~of hot SN shock front material could be incorporated into a cold molecular clump. But a 1 \msun~presolar cloud would need to be contaminated by $\sim$$10^{-4}$~\msun~of SN matter to explain the $^{26}$Al meteoritic abundance \citep{tak08}. Although these two estimates are close, the main issue lies in the short lifetime of a small molecular cloud embedded in a hot plasma: the lifetime of a 1-\msun~cloud against evaporation in the $>$10$^6$~K (Equation~(\ref{eq2})) superbubble interior is only $\sim$$10^5$~yr \citep{mck77}, much shorter than the duration of stellar main sequence. This scenario is therefore highly improbable. 

\section{$^{26}$Al production by a runaway WR star}

If the vast majority, if not all O-type stars (the main-sequence progenitors of WR stars) form in  clusters \citep[e.g.][]{lad03}, nearly half of them acquire  velocities exceeding the escape velocity from the cluster's potential well \citep{sto91}. These runaway stars\footnote{While traditionally the minimum peculiar velocity for classifying a star as a ``runaway'' is 40~km~s$^{-1}$ \citep{bla61}, here we use this term for any star that has escaped from its parent cluster \citep[see also][]{sto91}. The escape velocity ranges from several km~s$^{-1}$ for loose and low-mass clusters to several tens of km~s$^{-1}$ for compact and massive ones.} can be accelerated either by dynamical interactions with other stars in the dense cores of young clusters \citep{leo90} or by the SN explosion of a companion star in a massive binary system \citep{bla61}. A star moving with a velocity $V_* \gsim 15$~km~s$^{-1}$ relative to its parent cluster leaves the associated superbubble in less than 3~Myr (see Eq.~[\ref{eq1}]). About 20\% of the O-type stars have peculiar velocities exceeding 15~km~s$^{-1}$ \citep{dew05}. These runaway short-living stars may have a significant probability of interacting with their parent molecular cloud complex. Outside the hot gas, the star's motion is supersonic with respect to the ambient medium, which generates a bow shock \citep{van90}. There are many observations of bow shocks created by runaway OB stars in the vicinity of young clusters and associations \citep[e.g.][]{gva08}.

The form of a bow shock is determined by the balance between the ram pressure of the stellar wind and the ram pressure of the ongoing circumstellar (CS) gas. The pressure equilibrium is reached in the star's direction of motion at the so-called standoff distance from the star (\citealt{van90}; see Figure~1(b)),
\begin{equation}
R_0 = (0.92~{\rm pc}) \dot{M}_{W,-5}^{1/2} V_{W,1500}^{1/2}  n_{\rm H,100}^{-1/2} V_{*,20}^{-1}~,
\label{eq5}
\end{equation}
where $\dot{M}_{W,-5}$ is the stellar wind mass-loss rate in units of $10^{-5}$~\msun~yr$^{-1}$, $V_W=V_{W,1500} \times 1500$~km~s$^{-1}$ is the wind's terminal velocity, and $V_{*,20}=V_*/(20~{\rm km~s^{-1}})$. The interaction region between the stellar wind and the CS medium is a shell bounded by two shocks, in which the flows slow down from supersonic to subsonic velocities. One verifies that for $V_W \gg V_*$ the shell's mass is mainly due to the shocked CS gas. The contact surface between the shocked stellar wind and CS gas is unstable due to both the Kelvin-Helmholtz and Rayleigh-Taylor instabilities \citep{bri95}, such that we expect an efficient mixing of the wind-ejected material with the swept-up CS gas. 

\begin{figure}
\plotone{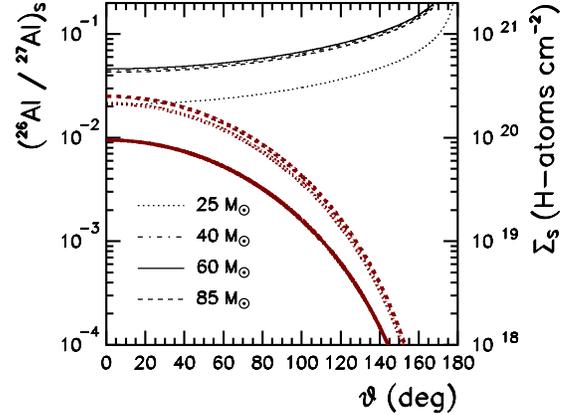}
\figcaption{Mean $^{26}$Al/$^{27}$Al ratio (thick curves, left axis) and H column density (thin curves, right axis) in the bow shock shell of a  runaway WR star, as a function of position $\theta$ in the shell (see Figure~1(b)), for $V_*=20$~km~s$^{-1}$ and $n_{\rm H}=100$~cm$^{-3}$. The $^{26}$Al/$^{27}$Al ratio is not shown for the 40~\msun~star (see the text); besides, the $\Sigma_S$ curve for this star merges with that for the 60 and 85~\msun~stars. 
\label{fig2}}
\end{figure}

The $^{26}$Al/$^{27}$Al ratio in the bow shock shell of a runaway WR star just prior to the SN explosion can be estimated as a function of the polar angle $\theta$ from the star's direction of motion (see Figure~1(b)) by 
\begin{equation}
\bigg({^{26}{\rm Al} \over ^{27}{\rm Al}}\bigg)_S(\theta) \simeq {N_{26}f_{26} \over 4 \pi x_{27} R_S^2(\theta) \Sigma_S(\theta)}~,
\label{eq6}
\end{equation}
where $N_{26}$ is the total number of $^{26}$Al nuclei ejected in the WR wind, $f_{26}=(\tau_{26}/\Delta_{\rm WR}) \times (1-\exp(-\Delta_{\rm WR}/\tau_{26}))$ is a factor that takes into account the decay of $^{26}$Al during the duration $\Delta_{\rm WR}$ of the WR phase ($\Delta_{\rm WR} \simeq 0.2$--1.4~Myr depending on the star's mass), $x_{27}=3.46 \times 10^{-6}$ is the $^{27}$Al abundance by number in the CS medium assumed to be of solar composition \citep{lod03}, $R_S(\theta)$ is the shell's radius (Figure~1(b)), and $\Sigma_S(\theta)$ is the shell's H column density ($\Sigma_S(0^\circ) \simeq 0.75R_0n_{\rm H}$; see \citealt{wil96}). We took the $^{26}$Al yields from the rotating stellar models of \citet{pal05}. $R_S(\theta)$ and $\Sigma_S(\theta)$ were calculated from the analytic solutions found by \citet{wil96} in the thin-shell approximation. The WR star parameters $\dot{M}_W$, $V_W$, and $\Delta_{\rm WR}$ were extracted from the grids of rotating stellar models of Meynet \& Maeder (2003; see also \citealt{vos09}).

Calculated $^{26}$Al/$^{27}$Al ratios in bow shock shells of runaway WR stars are shown in Figure~2 for stars of initial masses 25, 60, and 85~\msun~(the $^{26}$Al yield for the 40 \msun~star in not listed in \citealt{pal05}). The $^{26}$Al/$^{27}$Al ratio is only weakly dependent on the star's initial mass, because both $N_{26}$ and the amount of swept-up CS matter ($\propto R_S^2 \Sigma_S$) increase with increasing stellar mass. We see that ($^{26}$Al/$^{27}$Al)$_S$ reaches $\sim$1--2$\times$10$^{-2}$ at $0^\circ$. The isotopic ratio decreases with increasing $\theta$, because for the same solid angle as viewed from the star, the $^{26}$Al atoms ejected at backward angles are mixed with a higher mass of shocked CS gas. The mass contained in the shell's forward hemisphere (i.e., at $\theta < 90^\circ$) is nevertheless significant: it amounts to 23~\msun~for the 25 \msun~star and is between 200 and 250~\msun~for the three other stars. 

\begin{figure}
\plotone{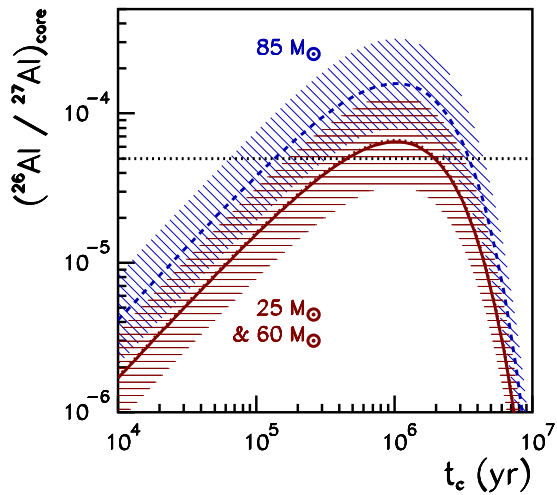}
\figcaption{$^{26}$Al/$^{27}$Al ratio in a prestellar core formed in the shocked-gas layer of a SN remnant resulting from the explosion of a runaway WR star, as a function of the time of core collapse after the explosion. Same as for Figure~2, $V_*=20$~km~s$^{-1}$ and $n_{\rm H}=100$~cm$^{-3}$. The hatched areas reflect a factor of 2 of uncertainty in the calculated isotopic ratio. The dotted horizontal line shows the canonical ratio measured in CAIs. The results for the 25 and 60~\msun~stars are almost identical. 
\label{fig3}}
\end{figure}

Hydrodynamic simulations of stellar wind bow shocks have shown that the steady-state solution of \citet{wil96} provides a good description of the time-averaged shape of the bow shock shell; although a bow shock is neither smooth nor steady \citep{rag97,blo98}. The shell is subject to periodic oscillations in and out with respect to the equilibrium position, which has been interpreted as resulting from the nonlinear thin shell instability (NTSI; \citealt{vis94}). This instability is also known to be an efficient dynamical focusing mechanism for large-scale gas stream, resulting in the buildup of dense cores. Thus, hydrodynamic simulations of the NTSI have shown that the density contrast in the shell can reach 10$^2$ to 10$^4$, depending on the gas cooling efficiency \citep[e.g.][]{hue03}. The high-density seeds thus generated are likely sites of further star formation \citep{hei08}. However, the required gravitational collapse of these dense cores is probably not possible as long as the shell is exposed to the intense photoionizing radiation from the nearby massive star \citep[see, e.g.,][]{rag97}. 

At the end of the WR stellar phase, the SN outburst will expel the bow shock material to large distances. About 10$^4$ years after the explosion, radiative cooling of the shock-heated gas will become important and the SN remnant will enter the pressure-driven snowplow phase. The transition from the adiabatic to the radiative phase in SN remnants is accompanied by the development of dynamical instabilities, that can further increase the mass of pre-existing gas clumps \citep{blo98b}. The timescale for collapse of a dense core embedded in a shocked gas layer is governed by the gravitational instability \citep{hei08} and reads \citep{whi94}
\begin{equation}
t_c \sim {2c_s \over 1.4m_{\rm H} G \Sigma_{\rm core}} \sim (2 \times 10^6~{\rm yr}) \bigg({\Sigma_{\rm core} \over 10^{22}~{\rm cm^{-2}}}\bigg)^{-1}~,
\label{eq7}
\end{equation}
where $\Sigma_{\rm core}$ is the H column density of the core, $c_s \approx 0.5$~km~s$^{-1}$ is the local sound speed, and $G$ is the gravitational constant. 

The $^{26}$Al/$^{27}$Al ratio in such a prestellar core can be estimated to be 
\begin{equation}
\bigg({^{26}{\rm Al} \over ^{27}{\rm Al}}\bigg)_{\rm core} \sim \bigg({^{26}{\rm Al} \over ^{27}{\rm Al}}\bigg)_S {\Sigma_S \over \Sigma_{\rm core}} \exp(-t_c/\tau_{26})~.
\label{eq8}
\end{equation}
This ratio is shown in Figure~3 as a function of $t_c$. We adopted for ($^{26}$Al/$^{27}$Al)$_S$ and $\Sigma_S$ the values at $\theta =45^\circ$ (see Figure~2). The error in ($^{26}$Al/$^{27}$Al)$_{\rm core}$ shown in Figure~3 is intended to account for various uncertainties in the model parameters, e.g., in $\theta$, $N_{26}$, and $c_s$. We see that ($^{26}$Al/$^{27}$Al)$_{\rm core}$ increases almost linearly up to $t_c \sim 10^6$~yr, thus reflecting that $\Sigma_{\rm core} \propto t_c^{-1}$ (Equation~(\ref{eq7})). The predicted $^{26}$Al/$^{27}$Al ratios are consistent with the canonical value measured in CAIs for a large interval of $t_c$ (as the delay for the CAI formation after collapse of the presolar nebula is $\ll \tau_{26}$, it can safely be neglected). The exponential decay of ($^{26}$Al/$^{27}$Al)$_{\rm core}$ for $t_c \gsim 10^6$~yr is due to the $^{26}$Al radioactivity. 

Inserting Equations~(\ref{eq5}) and (\ref{eq6}) into Equation~(\ref{eq8}) one can see that ($^{26}$Al/$^{27}$Al)$_{\rm core}$ scales as $n_{\rm H} \times V_*^2$. Thus, the solar system formation could have been triggered by the explosion of a runaway WR star propagating into atomic interstellar gas ($n_{\rm H} \approx 1$~cm$^{-3}$), but only if the star's velocity was $>100$~km~s$^{-1}$. 

In the proposed scenario, the incorporation of $^{26}$Al-rich stellar ejecta into interstellar gas is due to various dynamical instabilities operating in both the bow shock and the SN remnant shells. The associated turbulence is expected to homogenize the mixing at all scales, regardless of the carrier phase of $^{26}$Al (gas or dust). This is consistent with the Mg isotopic data of \citet{thr06} and \citet{vil09}, which suggest that $^{26}$Al was homogeneously distributed in the early solar system. 

\citet{arn06} showed that WR star nucleosynthesis can produce $^{26}$Al, $^{36}$Cl, and $^{41}$Ca at levels compatible with the meteoritic measurements, provided that the delay before the incorporation of these SLRs into CAIs was $\sim$1--$3 \times 10^5$~yr. But using the yields for a 60 \msun~star given by these authors, we obtain a $^{36}$Cl abundance well below the value reported in CAIs, as also found previously by \citet{gai09}. On the other hand, the present work shows that both $^{26}$Al and $^{41}$Ca abundances in meteorites can result from the contamination of the presolar molecular core by material ejected from a runaway WR star, whose explosion as a SN triggered the formation of the solar system. 

We thank Jean-Pierre Thibaud, Fa\"irouz Hammache, and Pierre Roussel for their critical reading of the manuscript.

\end{document}